\documentclass[12pt,preprint]{aastex}
\usepackage{graphics,graphicx}
\def \cm{~\rm{cm}}
\def \s{~\rm{s}}
\def \km{~\rm{km}}

\def \K{~\rm{K}}

\def \G{~\rm{G}}
\def \AU{~\rm{AU}}
\def \erg{~\rm{erg}}

\def \yr{~\rm{yr}}

\def \lesssim{\mathrel{<\kern-1.0em\lower0.9ex\hbox{$\sim$}}}
\def \gtrsim{\mathrel{>\kern-1.0em\lower0.9ex\hbox{$\sim$}}}

\begin{document}

\title{MAGNETIC FLARES ON ASYMPTOTIC GIANT BRANCH STARS}

\author{Noam Soker\altaffilmark{1,2} and Joel H. Kastner\altaffilmark{1}}

\altaffiltext{1}{Chester F. Carlson Center for Imaging
Science, Rochester Institute of Technology, 54 Lomb Memorial
Dr., Rochester, NY 14623; jhk@cis.rit.edu}
\altaffiltext{2}{Department of Physics, Oranim, Tivon 36006, 
ISRAEL; soker@physics.technion.ac.il}

\begin{abstract}
We investigate the consequences of magnetic flares on the
surface of asymptotic giant branch (AGB) and similar stars.
In contrast to the solar wind, in the winds of AGB stars the
gas cooling time is much shorter than the outflow time.  As
a result, we predict that energetic flaring will not
inhibit, and may even enhance, dust formation around AGB
stars.  {{{If magnetic flares do occur around such stars, we
expect some AGB stars to exhibit X-ray emission; indeed
certain systems including AGB stars, such as Mira, have been
detected in X-rays. However, in these cases, it is difficult
to distinguish between potential AGB star X-ray emission
and, e.g., X-ray emission from the vicinity of a binary
companion.  Analysis of an archival ROSAT X-ray spectrum of
the Mira system suggests an intrinsic X-ray luminosity
$\sim2\times10^{29}$ erg s$^{-1}$ and temperature $\sim10^7$
K. These modeling results suggest that magnetic activity,
either on the AGB star (Mira A) or on its nearby companion
(Mira B), is the source of the X-rays, but do not rule out
the possibility that the X-rays are generated by an
accretion disk around Mira B.}}}
\end{abstract}


\section{INTRODUCTION}

Asymptotic giant branch
(AGB) stars represent the final stages of evolution for
intermediate mass ($\sim1-8 M_\odot$) stars, just before these
stars (potentially) generate planetary nebulae. Such
stars have exhausted their core hydrogen, have
completed core helium and H-shell burning
phases, and are now burning H and He in concentric shells
around the exhausted stellar core (this core, once exposed, will 
become a degenerate white dwarf). AGB stars are ascending
fully convective Hayashi tracks in the H-R diagram, and the
very large luminosities of these
stars ($\sim10^4 L_\odot$), in combination with copious dust
grain formation in their cool, extended atmospheres, lead to
dusty, radiatively-driven winds. 
Hence, AGB stars lose mass rapidly, with rates ranging from
$\sim 10^{-8}$ to $\sim 10^{-4}$ $M_\odot$ yr$^{-1}$ (as
ascertained from far-IR continuum and
millimeter-wave molecular line emission; e.g. Loup et al.\ 1993).

There are several independent pieces of evidence for the
presence of magnetic fields in and around cool giants and,
in particular, in and around AGB stars.  These include: 
\begin{enumerate}
\item Maser polarization around AGB
stars and other cool stars. SiO masers are found close to
(typically within less than a few stellar radii of) the
stellar surface (e.g. Kemball \& Diamond 1997), OH masers
are found at at $\sim 10^{15}-10^{16} \cm$ from the star
(e.g., Szymczak, Cohen, \& Richards 1999; Palen \& Fix
2000; Bains et al. 2003), and H$_2$O masers are found in between
(Vlemmings, Diamond, \& van Langevelde 2002).  
\item Polarization of OH
maser spots in planetary nebulae (PNs) and proto-PNs, which
are the immediate descendants of AGB stars (e.g., Zijlstra
et al.\ 1989).  Miranda et al.\ (2001) find polarization in
the 1,665-MHz OH maser line, which indicates the presence of
$\sim 10^{-3} \G$ magnetic fields at $\sim 10^{16} \cm$ from
the central star of the young PN K3-35.  
\item X-ray emission
from cool giant stars (H\"unsch et al. 1998;
Schr\"oder, H\"unsch, \& Schmitt 1998; H\"unsch 2001).
H\"unsch (2001) used {\it Chandra} to detect X-ray emission
from two M giant stars, which display hard X-ray
luminosities of $L_x \simeq 10^{30} \erg \s^{-1}$.
\end{enumerate}

Although it is widely agreed that radiation pressure on dust is
the main acceleration process of winds from AGB stars   
(Elitzur, Ivezic, \& Vinkovic 2002, and references therein), 
there is still disagreement over the role the magnetic field plays
in shaping the circumstellar matter. 
First, it is not clear whether magnetic fields play any role
in shaping the winds; many other models exist, e.g., 
acceleration via radiation pressure on dust in fast rotating AGB stars
spun up by companions (e.g., Dorfi \& H\"ofner 1996;
Reimers, Dorfi \& H\"ofner 2000),  
and the influence of a companion outside the envelope
(e.g., Mastrodemos \& Morris 1999; Soker 2001). 
Second, there is a disagreement on whether the magnetic fields  
play a dynamical role, i.e., whether the magnetic
pressure and/or tension become comparable to the other relevant
forces in the flow  
(e.g., Pascoli 1997;
Chevalier \& Luo 1994; Garc\'{\i}a-Segura 1997;
Garc\'{\i}a-Segura et al.\ 1999; Garc\'{\i}a-Segura, \& L\'opez 2000;
Garc\'{\i}a-Segura, L\'opez, \& Franco 2001; 
Matt et al.\ 2000; Blackman et al.\ 2001; Gardiner \& Frank 2001;
Falceta-Goncalves \& Jatenco-Pereira 2002; 
see review by Garc\'{\i}a-Segura 2002), or whether the magnetic field
is relatively weak on a global scale, and plays only secondary effects
(Soker 2002; Soker \& Zoabi 2002 and references therein).

In earlier papers Soker and collaborators (e.g., Soker 1998; 
Soker \& Clayton 1999) propose that cool magnetic spots 
facilitate the formation of dust, hence locally increasing the 
mass loss rate. 
In this process the magnetic field has a secondary role and
becomes dynamically important only in relatively small
regions of the stellar surface and circumstellar environment. 
In this model the large scale magnetic field is too weak to play 
a dynamic role and directly influence the wind from the AGB star.
However, the magnetic field is strong enough above and near cool
spots that reconnection may occur {{{(if magnetic pressure exceeds
thermal pressure);}}} such reconnection events
likely would lead to magnetic flares analogous to those
on the surface of the sun. 
The reconnection process is basically cancellation of magnetic field
lines having opposite directions.
The energy released by the reconnecting magnetic fields heats 
the gas and accelerates it to high speeds.
The gas is heated by MHD shocks (e.g., Chen et al.\ 1999). 
In the present paper we discuss some implications of these events. 
In $\S 2$ the potential of impact of flaring on
circumstellar dust formation is considered. 
In $\S 3$ we discuss the expected observational signatures of
AGB star magnetic activity in general and of flares in particular.
Our summary is in $\S 4$. 

\section{DUST FORMATION IN FLARES}

Whereas Soker (1998) and Soker \& Clayton (1999) argued that
magnetic starspots on cool giants can lead to enhanced dust
formation, we demonstrate here that reconnection events
(flares) associated with such spots may themselves further
enhance the formation rate of dust. 
Because the density close to the surface of AGB stars is very high,
the question regarding dust formation reduces to the value 
of the temperature. 
To evaluate the temperature resulting from a flare, the time
scale of cooling should be compared with the flow time. 
In the sun the conduction time is non-negligible compared
with the radiative cooling time (e.g.,
{{{{ Vesecky, Antiochos, \& Underwood 1979; }}}}  
Aschwanden et al. 2000).
However, in AGB stars it can be neglected.
This can be shown by using equations (16) and (17) from Aschwanden
et al. 2000:
{{{{ taking a temperature of $T \sim 10^6 \K$ and electron
density of $n_e \sim 10^{10}$ in their equations (16) and (17)
yields a radiative cooling time of $\tau_{\rm rad} \sim 800 \s$
(see below), and a conduction time scale of
$\tau_{\rm cond} \sim  10^{9} (L/10^{12} \cm)^2 \s \gg \tau_{\rm rad}$,
where $L$ is the loop length.     }}}} 
 
Let us consider an idealized simple case, in which the
magnetic field expands radially from the stellar surface outward,
while the density is determined by an isothermal hydrostatic
halo. 
In such a case, the magnetic field decreases relatively slowly, 
$B(r)=B_\ast(r/R_\ast)^{-2}$, where $B_\ast$ is the magnetic field on
the stellar surface, and $R_\ast$ is the stellar radius
(subscript $\ast$ will stand for quantities on the stellar surface). 
For an isothermal gas above and close to the photosphere the density 
is given by 
\begin{eqnarray}
\rho=\rho_\ast \exp[-(r-R_\ast)/H_\ast],
\end{eqnarray}
where the scale height close to the stellar surface is 
\begin{eqnarray}
\frac{H}{R_\ast} \equiv \frac{C_\ast^2}{v_K^2} 
\simeq 0.045  
\left( \frac{T_\ast}{3000 \K} \right) 
\left( \frac{R_\ast}{1 \AU} \right) 
\left( \frac{M_\ast}{M_\odot} \right)^{-1}.
\end{eqnarray}
In equation (2), $C_\ast$ and $v_K$ are the isothermal sound
speed and Keplerian velocity on the stellar surface, respectively.
{{{The temperature was scaled for pre-flare gas because we do not
expect AGB stars to possess long lived coronae.
In any case, in calculating $C_\ast$ we took a fully ionized gas
so the equation can be used for higher temepratures. }}}
For reconnection to occur and heat the gas substantially,
the magnetic pressure should overwhelm the thermal pressure. 
This occurs several scale heights above the surface, and higher.
In the case of the sun the same condition is met very close
to the solar surface, and we can have small flares as well as large flares.
The smaller flares (nanoflares) above the solar surface have minimum height 
of $h \sim 0.0015 R_\odot \simeq 4 H_\odot$ (e.g., Aschwanden et al. 2000).
Inside magnetic flux loops the magnetic pressure is high, and
pressure equilibrium with the loop exterior implies that the density
inside the loop is smaller than in its surroundings. 
Also, after heating following reconnection, the hot gas expands
and the density drops below the simple isothermal halo value.
Strong flares extend to much larger distances from the solar surface,
$h \simeq 0.007-0.07 R_\odot \simeq 20-200 H_\odot$
(e.g., Aschwanden et al. 2000).

As can be seen from equation (2), on the surface of AGB stars,
for which $R_\ast \gtrsim 1 \AU$,
the scale height is $H \sim 0.05 R_\ast$. As just noted for
the case of the sun, flares, from small to large, extend from $h \sim 4 H$
to $h \sim 200 H$ above the solar surface. 
Taking the same range for AGB stars, for which
$H\simeq 0.05 R_\ast$ (Eq. 2), we find that flares will occur
over the range of $r=R_\ast + h \simeq 1.2-10 R_\ast$
Since in the upper range $h \gg R_\ast$, it is not
clear if flares can extend to such large radii
above the surfaces of AGB stars.
{{{{However, there is no physical reason why flares will not occur
on small scales, i.e., $h \sim 0.2-1 R_\ast$.
The magnetic energy can even be released in several consecutive
reconnection events, without changing our conclusions.
We also note that even in large flares the total magnetic energy
is still much below that required in
models where the magnetic field has a dynamical role (see Section 1).}}}} 

The typical density of the photospheres of AGB stars is 
$n_e(R_\ast) \sim 10^{14} \cm ^{-3}$ (Soker \& Harpaz
1999\footnote{Note that the density scale in Figs.\ 1-5 of
Soker \& Harpaz 1999
is too low by a factor of 10; the correct scale is displayed
in their Fig. 6.}).
In the simple isothermal halo used above, the density drops to
$n_e \simeq 10^{10} \cm^{-3}$ at $r \sim 1.3 R_\ast$.  
At distances of $r \gtrsim 2 R_\ast$ we are in the 
region where density is determined by the wind, i.e., 
$\rho = \dot M /(4 \pi r^2 v_w)$,
where $\dot M$ is the mass loss rate into the wind (defined positively),
and $v_w (r)$ is the wind speed.
For a fully ionized wind --- which is not the case for the
global AGB wind, but will be the case locally after a flare
event --- the scaled electron density is  
\begin{eqnarray}
n_e \simeq 10^{10} 
\left( \frac{\dot M}{10^{-6} M_\odot \yr^{-1}} \right) 
\left( \frac{r}{1 \AU} \right) ^{-2}
\left( \frac{v_w}{10 \km \s^{-1}} \right) ^{-1}
\cm^{-3} 
\end{eqnarray}
where, for simplicity, we have assumed isotropic mass loss.
Using the cooling function from, e.g., Gaetz, Edgar \& Chevalier (1988), 
we find the isobaric radiative cooling time in the temperature range 
$2 \times 10^5 K \lesssim T \lesssim 4 \times 10^7 \K$ to be 
\begin{eqnarray}
\tau_{\rm rad} \simeq 800 
\left( \frac {n_e} {10^{10}} \right)^{-1} 
\left( \frac {T} {10^6} \right)^{3/2} \s. 
\end{eqnarray}  

The flow time across the reconnection region can be estimated as follows.
Let the size of the reconnection flux tube be
$x \simeq H \sim 10^{12} \cm$, and the flow speed 
be the sound speed at $T = 10^6 \K$, i.e., 
$v_{\rm exp} \sim 100 \km \s^{-1}$.
The expansion time is then
\begin{eqnarray}
\tau_{\rm exp} \sim 10^5 
\left( \frac {x}{H} \right)
\left( \frac {H}{10^{12} \cm} \right) 
\left( \frac {v_e}{100 \km \s^{-1}} \right)^{-1} \s.
\end{eqnarray}
 Even if $x \sim 0.1 H$, the radiative cooling time of gas heated
by flares to $T \sim 10^6 \K$ in AGB stars is much shorter than the 
expansion time.
The cooling time will also be shorter than the expansion time
in strong and large flares, having a temperature of $T \sim 10^7 \K$
and a size of $x \gtrsim 10 H$. 
Close to the AGB stellar surface the wind does not yet reach its
terminal velocity, hence the density is higher and the cooling time
shorter than the values in equations (3) and (4), respectively. 
In the sun, where $H_\ast \sim 3 \times 10^7 \cm$, we find that the 
situation is different, with the radiative cooling time 
much longer than the expansion time (see, e.g., Aschwanden et al. 2000).
At larger distances from the surface of AGB stars, $r \sim 10 R_\ast$,
the cooling time is no longer much shorter than the expansion time;
nevertheless, the condition that the cooling time is much shorter
than the expansion time holds for dust formation close to the
stellar surface.

The compression in the strong shock that results from the
reconnection event can be larger than a factor of 4. This is
because the gas is not fully ionized.  For conditions above
cool giants, and for a shock temperature of $\sim 10^6 \K$,
the post shock density can be a factor $\sim 5-10 $ larger
than the pre-shock density (Woitke, Gores \& Sedlmayr
1996a).  Since the gas cools much faster than expansion
time, to maintain pressure equilibrium the gas will be
further compressed.  The situation here is more complicated
than the shocks generated by pulsating stars, as discussed by
Woitke et al.\ (1996a), because the shocks are MHD shocks
(e.g., Chen et al.\ 1999).  After the gas cools from $T \sim
10^6$ to $T \sim 10^4 \K$, we can take, therefore, a density
that is larger by a factor of $\sim 10$ relative to the value in
equation (3).

After this gas cools to $T \simeq 10^4 \K$, the radiative cooling time
increases substantially (e.g., Woitke, Kr\"uger, \& Sedlmayr 1996b);
on the other hand, the cooling now proceeds radiatively
as well as via adiabatic expansion (Woitke et al.\ 1996a). 
To form dust, the temperature should decrease to $T \lesssim 10^3 \K$.
Using figure (11) of Woitke et al.\ (1996b), we approximate the 
cooling time from a temperature of $T\simeq 10^4 \K$ down to 
dust-forming temperatures by
\begin{eqnarray}
\tau_{\rm rad} \simeq 5 \times 10^6 
\left( \frac {n_H}{10^{10} \cm^{-3}} \right)^{-0.3} \s,
\end{eqnarray}
{{{{ where $n_H$ is the hydrogen number density.
Note that the gas is not fully ionized in this case, in
contrast to the temperature range for which equation (4) holds.
The cooling time is still short, however. }}}} 
The cooler gas will expand at slower velocity; either the sound speed or
the wind speed give $v_e \simeq 10 \km \s^{-1}$, which gives for the
expansion time $\tau_e \sim 10^6 \s$, according to equation (5).
Taking $n_H \sim 10^{11} \cm^{-3}$, as discussed above, and 
considering further cooling by adiabatic expansion (Woitke et al.\ 1996a), 
we find the expansion time of the gas to be comparable to the cooling time
for the scaling used above, i.e., a mass loss rate 
of $\dot M = 10^{-6} M_\odot \yr^{-1}$ in equation (3).

The main conclusion from this discussion is that magnetic flares
on the surfaces of AGB stars, as well as on other classes of actively
mass-losing cool giant stars
{{{{ which lose mass at a high rate, }}}} 
{{{ will not inhibit dust formation, because
the gas cools on a short time scale.
The basic mechanism is that the lower temperature above cool
magnetic spots allows dust to form closer to the AGB surface
where the density is higher, thereby enhancing dust formation
and, consequently, mass loss rate (Soker 1998, 2000; Soker \& Clayton 1999).
{{{{ Furthermore, strong shocks, as expected in the reconnecting region,
may compress the gas such that the final temperature is
lower than that of the ambient medium. As the entire
cooling cycle occurs on a time  
scale during which the gas is still close to the star, the
density is relatively high.
Such a cooling cycle would then be similar to
that proposed for R Coronae Borealis (RCB) stars (Woitke et
al. 1996a). Evidently dust can be formed close to the
surfaces of RCB stars, although their photospheres are much
hotter than those of AGB stars.
In RCB stars, dust formation is
enhanced via shocks due to stellar pulsations. 
If flaring results in similarly strong shocks, the high
densities close to the AGB surface, e.g., at radii of
$\sim 1.2 R_\ast \lesssim r \lesssim 3 R_\ast$, could facilitate dust
formation. We therefore suggest that flaring may enhance
dust formation, }}}} 
as long as the mass loss rate is $\dot M \gtrsim 10^{-6} M_\odot \yr^{-1}$.
The condition may be somewhat different for very massive cool giants,
which we do not consider here, although the same qualitative mechanism
may apply. 

\section{OBSERVATIONAL SIGNATURES OF AGB STAR MAGNETIC ACTIVITY}

As the preceding suggests, it is likely that AGB stars
generate locally strong surface magnetic fields, resulting
in violent reconnection events (flares). Establishing the
presence of such transient magnetic activity on AGB stars
observationally would have several important ramifications;
in particular, we would gain new 
insights into stellar dynamos and the shaping of planetary
nebulae.  We now discuss several new observational
directions that show promise for establishing the degree of
magnetic activity around AGB stars.

\subsection{X-ray emission and flaring}

Magnetic activity --- due either to differential rotation in
convective regions or, for very young systems, to
interactions between magnetic fields associated with star and
circumstellar disk --- is widely believed to be ultimately responsible for
X-ray emission from late-type stars.
The physical conditions that give rise to this activity, and the
connection between, e.g., stellar magnetic dynamo activity and X-ray
emission, remains uncertain.
 It is apparent, however, that highly convective late-type stars and
young, low-mass stars that presumably retain high rotation velocities
and/or accretion disks are among the strongest X-ray emission sources.
Most low-mass, pre-main sequence (T Tauri) stars,
for example, have X-ray luminosities (relative to
bolometric) exceeding that of the Sun by 3-4 orders of
magnitude (Feigelson \& Montmerle 1999). Hence, if AGB stars
indeed generate strong magnetic fields, either locally or
globally, then {{{they may be}}} be sources of X-ray
emission and, perhaps, strong, long-duration (see below) X-ray flares.

\subsubsection{X-ray emission from the Mira system}

{{{ In addition to the apparent X-ray detection of two first-ascent red
giants (H\"unsch 2001), there exist two potential
X-ray detections of AGB stars thus far. Both stars are
in binary systems; one is a prototypical object,
Mira, and the other is the symbiotic system R Aqr (Jura \&
Helfand 1984). The X-ray emission from R Aqr was recently
imaged with the Chandra X-ray Observatory by Kellogg et al.\
(2001) and appears to originate in a disk/jet
system. Karovska et al.\ (1996) attribute the X-ray emission
from the Mira system to accretion onto Mira's companion,
Mira B. Other observations also suggest the existence of an
accretion disk around Mira B (e.g., Reimers \& Cassatella
1985; Bochanski \& Sion 2001, Wood Karovska \& Raymond
2002); Jura \& Helfand (1984) and Wood et al.\ (2002) note
that it is not clear if this companion is a white dwarf or a
main sequence star.

Here, we re-examine ROSAT X-ray observations\footnote{These ROSAT data were
obtained through the the High Energy Astrophysics Science
Archive Research Center, a service of the Laboratory for
High Energy Astrophysics at NASA/GSFC and the High Energy
Astrophysics Division of the Smithsonian Astrophysical
Observatory.} of Mira ($o$ Ceti), in the
context of the model presented in \S 2.
At a distance of 128 pc, Mira is one of the closest AGB stars to
Earth. ROSAT detected X-ray emission toward Mira during a 9144 s
pointed Position-Sensitive Proportional Counter (PSPC) observation on
1993 July 15 (Karovska et al.\ 1996). The 
background-subtracted PSPC count rate was 0.007
s$^{-1}$. Since only $\sim63$ counts are attributed to the source
($\sim10$ counts are background), the archival PSPC spectrum
of Mira is very noisy. Nevertheless, these data provide constraints on
the nature of the X-ray source. Fitting an absorbed
Raymond-Smith (RS) plasma model to the background-subtracted
spectrum (Fig.\ 1), we find the best-fit X-ray
emission temperature and absorbing column are $T_x =
1.0\times10^7$ K and $N_H = 1.8\times10^{21}$ cm$^{-2}$,
respectively (with formal uncertainties of $\pm10$\% in
$T_x$ and $\pm30$\% in $N_H$), with a total (absorbed) model
flux of $6\times10^{-14}$ ergs cm$^{-2}$ s$^{-1}$ (0.2-2.2
keV).  The inferred intrinsic X-ray luminosity of the RS
model is $\sim2\times10^{29}$ erg s$^{-1}$, consistent with
the estimate obtained by Jura \& Helfand (1984). An
absorbed blackbody --- a model perhaps more
appropriate to describe accretion-generated X-ray emission
from a disk around Mira B --- also produces acceptable
fits; however, the temperature 
required for an adequate fit ($\sim9\times10^5$ K) is an order of
magnitude higher than inferred via 
UV observations (Reimers \& Cassatella 1985), and the inferred
effective blackbody radius is only $\sim10^6$ cm. Furthermore,
the value of $N_H$ required to produce an adequate fit to
the PSPC data is very large 
($1.5\pm0.5\times10^{22}$ cm$^{-2}$). In contrast,
the estimate for absorbing column obtained by fitting the
RS model, $\sim2\times10^{21}$ cm$^{-2}$, is consistent with
the modest mass-loss rate of Mira as deduced
from circumstellar CO radio line emission ($\sim10^{-7}$
$M_\odot$ yr$^{-1}$; e.g., Loup et al.\ 1993),
although it is somewhat larger than the H {\sc i}
column density deduced from ultraviolet observations
of H$_2$ fluorescence (Wood et al. 2002).

These results suggest that the X-ray emission is coronal
in nature and, therefore, that magnetic activity is
responsible for the emission. There are many
possibilities as to the source(s) of this activity. It may occur
on Mira A itself, or on Mira B if it is a
low-mass main-sequence star (as suggested by Jura \& Helfand
1984; see also Soker \& Kastner 2002). Alternatively, the
emission may originate from B's disk, either in star-disk
magnetic field reconnection 
events or in a disk corona.}}} {{{{ Below, we further consider the
implications of the first possibility, i.e., X-ray emission
from Mira A.}}}} 

\subsubsection{X-ray properties of AGB star flares}

{{{{ If the X-ray emission detected by Einstein and ROSAT
originates in magnetic flaring on Mira 
A, then the ROSAT results (\S 3.1) suggest fundamental differences
between AGB star and solar flares. 
Using the X-ray luminosity quoted above then, assuming $n_e=10^{10} \cm^{-3}$
and a cooling function of
$\Lambda = 5 \times 10^{-23} \erg \cm^3 \s^{-1}$ at
$T \simeq 10^7 \K$, (Gaetz et al. 1988), we find the
emitting volume to be $V_{\rm em} = 5 \times 10^{31} \cm^{3}$.
For a loop with scale length $10^{12} \cm$ and cross section
$(10^{11} \cm)^2$, the filling factor of the emitting gas is
$\epsilon \simeq 5 \times 10^{-3}$.
That is, only a very small fraction of the
loop is responsible for the observed X-ray emission.
The filling factor can become close to unity if the electron
density is lower, $n_e \sim 10^9 \cm^{-3}$.
Such a low density is reasonable for Mira A, given its
relatively low mass-loss rate. Even for this lower density
the radiative cooling time is somewhat shorter than the expansion time, by
equations (4) and (5), if the loop 
is smaller than the scaling in equation (5).
Therefore, to hold the approximate analogy to solar flares,
then the X-ray emitting flares on Mira appear to be of
lower density than solar.
It is assumed also that the emitting gas resides in the magnetic
flux loop prior to reconnection, and there is no need for it to flow
from the photopshere to a distance of $\sim 10^{12} \cm$ as
the flare starts. Here, indeed, the strongly pulsating atmosphere of
AGB stars is markedly different from the solar atmosphere. }}}} 

In \S 2 we found that magnetic flares close to the
the surfaces of AGB stars, and other cool giant stars, have scales of
$\gtrsim 0.2 R_\ast \sim 5 \times 10^{12} \cm$. 
If magnetic field reconnection proceeds at the sound speed of the gas 
heated to $\gtrsim 10^6 \K$, which is some fraction of the Alfven speed 
prior to reconnection, then the duration of a typical flare is 
$\tau_{\rm flare} \gtrsim 5~$days. 
A long flare might last for of order a year.
This is in contrast to flares on solar-like stars, which have typical
time scales of $1-10^4 \s$ (e.g., Pettersen 1989). 
Another source of flares, consisting of transient, bright
emission from the optical through the X-ray regimes,
can be accretion events onto close companions, including brown dwarfs
and massive planets (Struck, Cohanim \& Willson 2002).
(The latter mechanism likely can't account for the X-ray properties of the
Mira system, however, as X-ray emission generated via
accretion onto a low-mass companion would be softer than that observed.)  
As with magnetic flares on main sequence stars and brown dwarfs,
the typical flare timescales of such events should be very
short compared with  
those expected from magnetic flares on AGB stars. It is
clear, therefore, that long-duration X-ray observations of AGB
stars are very desirable as a means of establishing the
presence and timescales of flares and, hence, distinguishing between
potential flare sources.

We also concluded in the last section that for high enough mass loss rate,
typically $\dot M \gtrsim 10^{-6} M_\odot \yr^{-1}$, flares may enhance
dust formation; in Mira, for example, the mass loss rate is
too low for flare-enhanced dust formation, even if the X-ray emission
originates with magnetic activity on the AGB star.
The magnetic fields enhance dust formation mainly via the formation
of cool spots (e.g., Soker 1998; Soker \& Clayton 1999).
Once local dust formation occurs, further dust formation can occur
because of shadowing of the stellar radiation (Woitke, Sedlmayr, \&
Lopez 2000; Soker 2000), hence further increasing the inhomogeneity of
the wind. Thus, in addition to searching for evidence for
flaring via X-ray observations of AGB stars, it would be
desirable to conduct coordinated X-ray and infrared
observing campaigns designed to detect enhanced dust
formation episodes during flares.

\subsection{Polarized radio-line maser emission: magnetic
clouds in AGB winds?}

Estimates of magnetic field intensities from polarization
in maser spots around AGB stars give high values 
(e.g., Vlemmings et al.\  2002). 
Although the exact estimated values of the magnetic field are 
uncertain (Elitzur 1996), the estimates are still accurate to 
an order of magnitude (Elitzur, M., private communication, 2002),
and in many maser spots implies that the magnetic pressure is 
much larger than the thermal pressure (Vlemmings et al.\ 2002;
Bains et al. 2003).
In the solar wind, on average, the magnetic pressure does not
exceed the thermal pressure. 
However, the magnetic pressure does exceed the thermal pressure in 
magnetic clouds (e.g., Burlaga 2001; Yurchyshyn et al.\ 2001).
Magnetic clouds are formed by impulsive mass loss events from
the sun, and they are characterized by stronger than average 
magnetic fields, low proton temperatures, and smooth rotation of the
magnetic field direction (e.g., Burlaga 2001). 
We suggest that the maser spots with strong magnetic fields
which are observed around some AGB stars are similar in nature
to the magnetic clouds in the solar wind, in that they
represent local enhancements of the magnetic field (Soker 2002;
see also Dorch \& Freytag 2002 for a possible local dynamo in Betelgeuse).
On average, we expect the magnetic pressure in the wind
of AGB stars to be much below the thermal pressure, as
winds from AGB stars are driven by radiation pressure on dust,
rather than magnetic activity as in the sun.

\section{SUMMARY}

Recently, evidence has accumulated for the presence of
relatively strong magnetic fields around AGB and similar stars.
We have investigated certain theoretical and observational consequences
of magnetic activity in such cool giant stars. The main
results of this investigation are as follows.
\begin{enumerate}
\item In contrast to the solar wind, in the expanding circumstellar
envelopes of AGB stars the post-flare gas cooling time is much shorter
than the typical flow timescale.
As a result, we predict that energetic flaring will not inhibit
dust formation around AGB stars, {{{ and may even enhance dust formation
in some cases. }}}
\item Magnetic reconnection events near the
stellar surface should lead to localized, long duration
(timescales $\sim$ few days to a year) flares. 
{{{ 
\item X-ray observations should provide indications as to whether
AGB stars display magnetic flare activity. 
Analysis of an archival ROSAT X-ray spectrum of the Mira system
yields results for intrinsic 
X-ray luminosity ($\sim2\times10^{29}$ erg s$^{-1}$) and
temperature ($\sim10^7$ K) that are 
consistent with such activity, either on Mira A or its
companion, Mira B, although we do not rule out
the possibility that the X-ray emission arises from an
accretion disk around Mira B.
}}}{{{{ X-ray spectroscopic
observations with XMM-Newton, whose sensitivity and spectral
resolution far exceed those of ROSAT/PSPC, should better
constrain the X-ray emission mechanism of the Mira system.
Meanwhile, the exceptional spatial resolution of the Chandra
X-ray Observatory likely provides the only means to
determine unambiguously the source of the X-ray
emission. The separation between Mira A and B, $\sim0.6''$
(Karovska et al.\ 1997), is just at the limit of Chandra's
spatial resolving power and absolute astrometry.  }}}}
\item Observations of polarized maser emission from
the inner circumstellar envelopes of AGB stars may indicate
the presence of localized, highly magnetized wind clumps ---
analogous to magnetic clouds in the solar wind --- rather
than large-magnitude global magnetic fields.
\end{enumerate}

\acknowledgements{
{{{ We thank an anonymous referee for comments which led us
to clarify many points. }}}
We acknowledge support for this research
provided by NASA/CXO grants GO0--1067X and GO2--3009X to RIT. 
N.S. acknowledges support from the US-Israel
Binational Science Foundation and the Israel Science Foundation. }

\begin{figure}[htb]
\includegraphics[scale=1.,angle=0]
{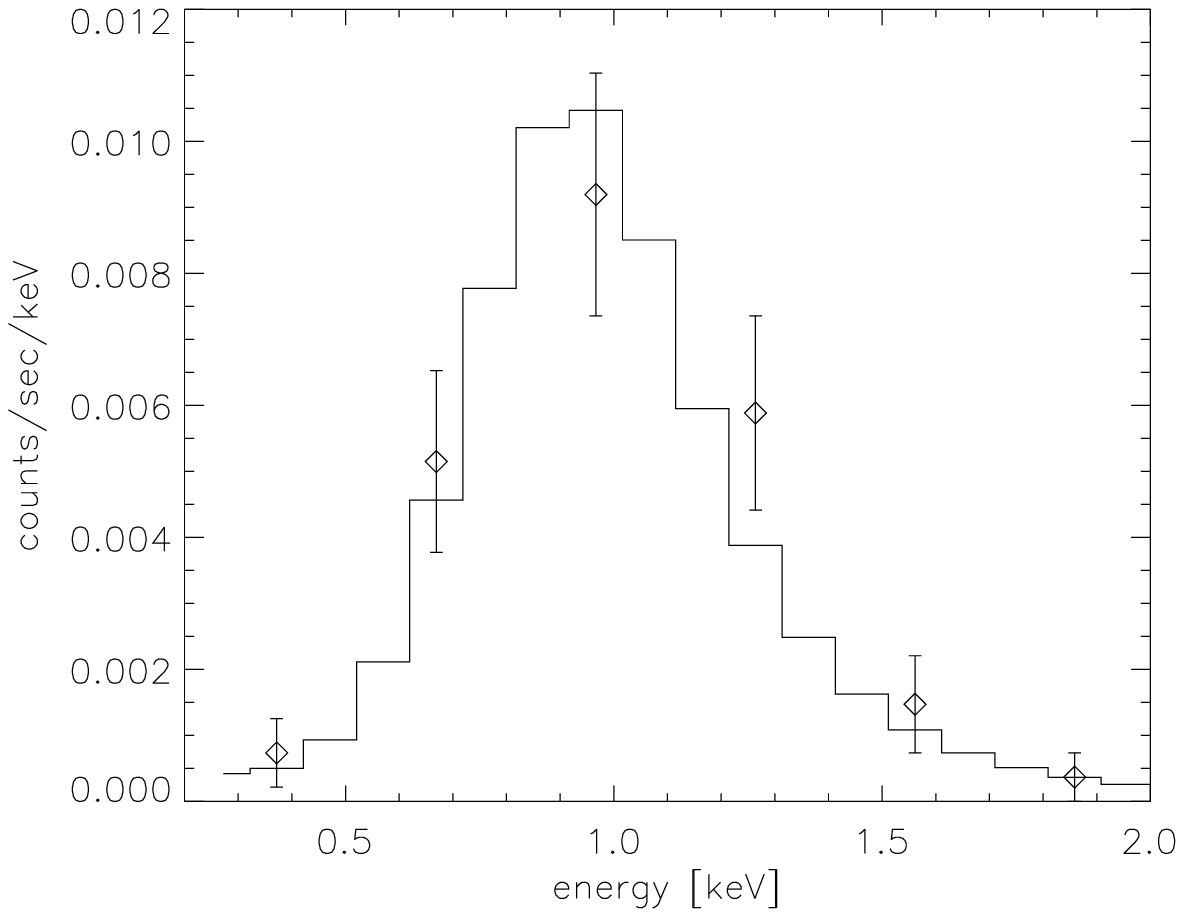} 
\caption{Archival ROSAT PSPC spectrum of the Mira system (diamonds),
overlaid with best-fit absorbed Raymond-Smith plasma model
(histogram). Error bars are 1 $\sigma$.} 
\end{figure}

\end{document}